\documentclass[doublecol]{epl2} 
\title{Robustness and Assortativity for Diffusion-like Processes in Scale-free Networks}
\author{G. D'Agostino\inst{1} \and A. Scala\inst{2,3} \and V. Zlati\'c\inst{4} \and G. Caldarelli\inst{5,3}}
\shortauthor{G. D'Agostino \etal}

\institute{                    
	\inst{1} ENEA - CR "Casaccia" - Via Anguillarese 301 00123, Roma - Italy\\
	\inst{2} CNR-ISC and Department of Physics, University of Rome ``Sapienza"
	P.le Aldo Moro 5 00185 Rome, Italy\\
	\inst{3} London Institute of Mathematical Sciences, 22 South Audley St
	Mayfair London W1K 2NY, UK\\ 
	\inst{4} Theoretical Physics Division, Rudjer Bo\v{s}kovi\'{c} Institute,
	P.O.Box 180, HR-10002 Zagreb, Croatia\\
	\inst{5} IMT Lucca Institute for Advanced Studies,
	Piazza S. Ponziano 6, Lucca, 55100, Italy\\
}
\pacs{89.75.Hc}{Networks and genealogical trees}
\pacs{05.70.Ln}{Nonequilibrium and irreversible thermodynamics}
\pacs{87.23.Ge}{Dynamics of social systems}

\abstract{
By analysing the diffusive dynamics of epidemics and of distress in complex networks, 
we study the effect of the assortativity on the robustness of the networks. 
We first determine by spectral analysis the thresholds above which
epidemics/failures can spread; we then calculate the slowest diffusional times.  
Our results shows that  disassortative networks exhibit a higher epidemiological threshold and are therefore easier to immunize, while in assortative networks there is a longer time for  intervention before epidemic/failure spreads. 
Moreover, we study by computer simulations the sandpile cascade model, a diffusive model of distress propagation (financial contagion). We show that, while assortative networks are more prone to the propagation of epidemic/failures, degree-targeted immunization policies increases their  resilience to systemic risk.
}

\begin{document}

\maketitle

\section{Introduction}

The heterogeneity in the distribution of contacts in a population 
is one of the key factors affecting the propagation of diseases \cite{PSatorrasPRL01}.
For example, a large variance of the degree (the number of a node neighbours) distribution 
is a typical feature of complex networks \cite{BAreview,CaldarelliBOOK07} that plays a role in 
determining the dynamical process defined on the networks itself \cite{BuldyrevNAT10}. 
It has been shown \cite{BogunaPRL03,DurrettPNAS10} that in the presence of a large 
heterogeneity,  the value of epidemic threshold tends to vanish in the limit of 
infinitely large network  
leading therefore to a finite probability of pandemic 
outbreak \cite{EubankASC10,YinPS07}.  
Such results have been obtained for specific models of diffusion processes as SIR and SIS on 
complex networks \cite{PSatorrasPRL01,DurrettPNAS10,CastellanoPRL10}.
Similar considerations apply to the analysis of any kind of propagation, as that of financial distress, even if in 
the latter case the channels of propagation are different from the epidemiological ones. 
Most of the derivation of the analytical results are based on 
mean field hypothesis and on analysis of the spectral properties of suitable 
matrices associated with the network 
\cite{BogunaPRE02,Wang03,Chakrabarti08,DurrettPNAS10} 
(we describe that matrices in detail in the following). 
This kind of studies is 
particular useful in order to define suitable procedures to stop the propagation of 
an epidemic \cite{MorenoPRE03,BadhamTPB10,KissJTRSI08}. 
 
In comparison to the humongous efforts that have been devoted to understand the role of 
the distribution of contacts in the networks, less attention  
has been paid to the assortativity (or vertex-vertex degree correlation) of the networks.
Actually, most if not all real networks have non-trivial values of this 
vertex-vertex correlation. In particular, some networks exhibit ``assortative mixing'' 
on their degrees, i.e. high-degree vertices tend to be attached to high-degree vertices; other networks show "disassortative mixing",  
i.e. high-degree vertices tend to be attached to small-degree vertices. 
The network's degree--degree
correlation can be summarized by a single scalar quantity $r$ called
the assortativity coefficient \cite{NewmanPRL02}. This quantity assumes
the value $r=0$ for degree-uncorrelated networks, is $r>0$ for
assortative networks and $r<0$ for disassortative ones. Assortative correlations are typically observed in social networks \cite{NewmanPRL02}, while disassortative connections are mainly found in
technological and biological networks \cite{NewmanPRE03}. 

We produce and analyse different networks with the same degree sequence but different assortativity. 
We first focus on spectral analysis, a very general approach to determine the 
diffusion dynamics on a complex network by matrix analysis; in particular, we derive the difference in epidemics propagation for networks with different assortativity properties.
We then focus on a classical model of distress propagation, the BTW sandpile\cite{BTW87}; 
using computer simulations, we study the effectiveness of targeted immunization policies
with respect to the assortativity of the underlying network. 

\section{Methods}

Formally, a network (or a graph) is defined as a pair $G=\left(V,E\right)$
where $V$ is the set of $N_{V}$ nodes and $E$ is the set of $N_{E}$
links; each link joins two nodes. To each graph $G$ we associate
its adjacency matrix $A$, defined as $A_{ij}=1$ if nodes $i$,$j$
are connected, $A_{ij}=0$ otherwise. We consider networks that are simple  
(no self loops, i.e. $A_{ii}=0$) and undirected ($A_{ij}=A_{ji}$).
The degree of node $i$ is the number of its neighbours and is 
defined as $k_{i}=\sum_{j}A_{ij}$.
The Laplacian matrix of a network is defined as $\mathcal{L}\equiv K-A$, where
$K$ is the diagonal matrix of degrees $K_{ij}=k_{i}\delta_{ij}$.
The Laplacian matrix is the analogous of the Laplacian operator and
describes the diffusion of random walkers on the network.

The assortative coefficient $r$ is the degree-degree Pearson correlation coefficient of two vertices 
connected by an edge 
$r= 
[\langle kq\rangle - \langle (1/2)(k+q)\rangle^2]/
[\langle (1/2)(k^2+q^2)\rangle - \langle (1/2)(k+q)\rangle^2]
$ 
where $k,q$ are the degrees of two adjacent vertices;  averages $\langle \ldots\rangle$ 
are over the edges. In order to sample the space of possible networks 
with respect to the assortativity, we use the procedure of 
ref.\cite{NohPRE07}. 

We sample the statistical ensemble of graphs $\lbrace G \rbrace$ with probability measure 
$\mu(G)\propto e^{-H(G)}$ induced by the Hamiltonian
$H\left(G\right)=-J{\displaystyle \sum_{ij}A_{ij}k_{i}k_{j}}$, where J 
is the coupling constant measuring the strength of the interaction.

\begin{figure}
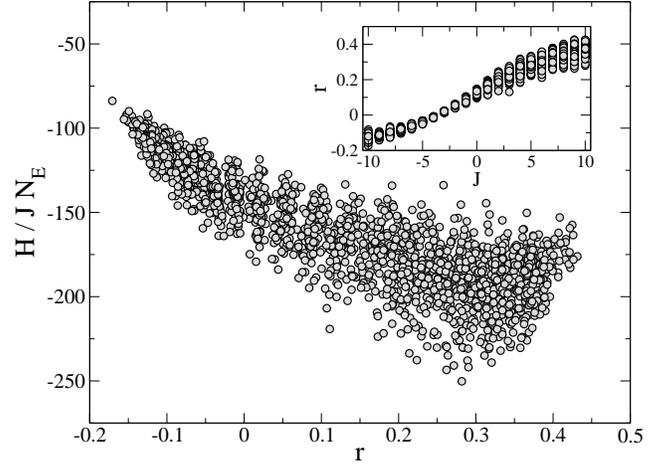

\onefigure[width=0.95\columnwidth]{fig1.eps}
\caption{Values of the energy per link $H/JN_{E}$, versus the assortativity
coefficient (i.e. degree-degree correlation) $r$, for all the $2100$ networks of $10000$ nodes.
In the inset the coefficient $r$ for the same networks as a function of $J$.}
\label{fig1}
\end{figure}

In order to sample configurations according to $\mu(G)$, we explore the configuration 
space by link rewiring \cite{MaslovSCIENCE02}
and accept a link rewiring with probability 
$e^{-\Delta H}$. Once the initial network is given, such procedure leaves 
the initial degree sequence unchanged.
We explore $21$ equally spaced values of $J$ ranging from $-10$ to $10$
producing $100$ independent configurations for each $J$.
Since $H\propto\langle k_i k_j \rangle$, 
the average $H/J$ sampled in this statistical ensemble 
decreases if the assortativity increases and {\em vice-versa}
(Fig.\ref{fig1}). 
Notice that the values of the assortativity $r$ with respect 
to the parameter $J$ is monotonously increasing (Fig. \ref{fig1}, inset).

\section{Spectral Analysis}

While the sampling procedure is general, in this paper we focus
on initial network configurations obtained by the Barab\'asi-Albert 
preferential attachment (BA networks) \cite{BarabasiSCI99}. 
For each value of $J$, we average over $10^2$ networks of $10^4$ nodes each. 

We first calculate the maximum eigenvalue $\Lambda_{1}$ of the adjacency matrices. 
The adjacency matrix $A$ dictates which nodes can be immediately 
reached by a "virulent" node and is hence central in describing not only the 
propagation of epidemics but also the propagation of faults/failures \cite{Wang03}; 
its maximum eigenvalue $\Lambda_{1}$ is related to how fast a process can spread
in a network.

For the SIS model of infections, Wang and coauthors
have shown that the epidemic threshold $\tau$ of a network is exactly 
$\tau=\Lambda_{1}^{-1}$ \cite{Wang03}; 
$\tau$ is defined  as the critical ratio among the propagation rate 
and the recovery rate of a disease above which epidemics ensue.

\begin{figure}
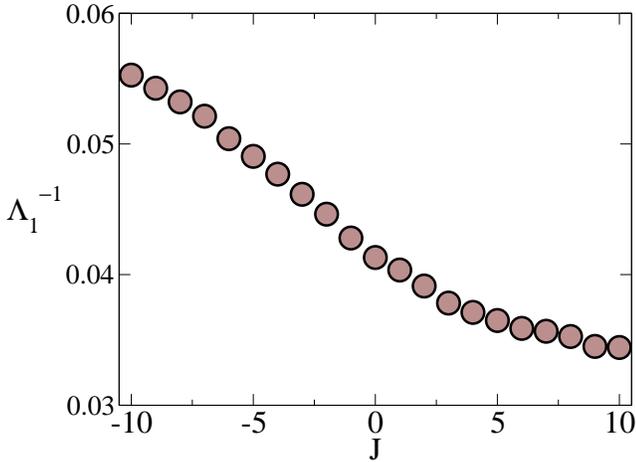

\onefigure[width=0.95\columnwidth]{fig2.eps}
\caption{$\Lambda_1^{-1}$ versus $J$. 
The decrease of $\Lambda_{1}^{-1}$ with $J$ and therefore with 
assortativity indicates a lower epidemic threshold; hence disassortative 
networks are less prone to epidemic spreading.
}
\label{fig2} 
\end{figure}

We find that $\Lambda_{1}^{-1}$ decreases with assortativity: in
the range of correlation explored, the most disassortative networks show
an epidemic threshold about $60\%$ higher than most assortative 
ones (Fig.\ref{fig2}). 
Our findings confirm the idea that avoiding direct connections between
hubs (highly connected nodes) may provide protection against
epidemics \cite{EguiluzPRL02}.

We then calculate the first non-zero eigenvalue $\lambda_{2}$ 
of the Laplacian matrices.
The Laplacian $\mathcal{L}$ describes the diffusive process 
$ \partial_t \mathbf{\rho} = -\mathcal{L}\mathbf{\rho} $ 
on the network. While the first eigenvalue $\lambda_{1}=0$ is associated to
the stationary distribution, the first non-zero eigenvalue $\lambda_{2}$ is 
the inverse timescale of slowest mode of diffusion. 

\begin{figure}
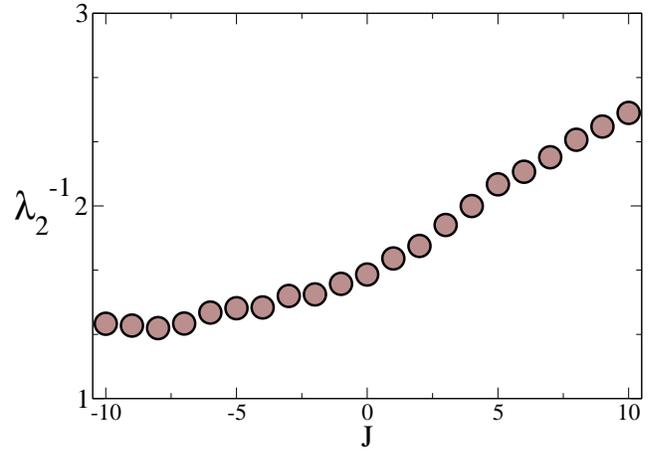

\onefigure[width=0.95\columnwidth]{fig3.eps}
\caption{$\lambda_2^{-1}$ versus $J$.  
The increase of $\lambda_{2}^{-1}$ with $J$ and therefore with 
assortativity indicates a growth of the longest relaxation time for 
diffusion processes on the network. Analogously, both the slowest 
vibrational mode and the synchronization time increase with 
assortativity.
}
\label{fig3} 
\end{figure}

In general, we can think of 
$\lambda_{2}^{-1}$ 
as the longest timescale after which a perturbation 
(like the infection of a site) that spreads diffusively  will settle 
a new state (like an epidemics) in the network. Therefore, a small 
value of $\lambda_{2}^{-1}$ means that there is less time for 
intervention before a network is totally compromised by randomly 
propagating failures or epidemics; in this respect most assortative networks 
show times up to $80\%$ higher than most disassortative 
ones (Fig. \ref{fig3}).

The eigenvalues of $\mathcal{L}$ govern also the harmonic dynamics of a network:
the adimensional vibration equation in the node displacements $x$ can in fact be written as 
$\partial^2_t \mathbf{x} = -{\displaystyle \sum_{j}A_{ij}(x_i-x_j) } = -\mathcal{L}\mathbf{x}$ 
and again the period of the slowest oscillatory 
mode is $\lambda_{2}^{-1}$. Almendral and coauthors have shown 
\cite{AlmendralNJP07} that in general synchronization times in complex 
networks have an almost linear dependence on $\lambda_{2}^{-1}$.

\section{Distress propagation}

Beside epidemiological models, there are other types of contagion propagation 
paradigms that are especially suitable for accounting cascades of financial distress.
Avalanche dynamics or {\em domino effect} (as it is frequently
indicated in economics) is currently regarded as an important feature during
financial crises. 
These phenomena are frequently modelled by means of a 
class of cellular automata known as sandpile models \cite{BTW87} that can mimic financial
distress propagation\cite{FeigenbaumRPP03};
financial crisis fluctuations are described, at least qualitatively, by the avalanches 
of a sandpile model\cite{Bartolozzi06}. 
Conservation laws provide the connection between sandpile models and 
random-walk (diffusive) models \cite{ShiloPRE03}. 

In sandpiles 
every vertex has a given capacity to store a scalar field (originally ``sand'').
For our purposes, such a scalar can represent financial distress (debts) or the
probability of a failure. When such a quantity reaches a threshold value, the
vertex becomes ``bankrupt'' and passes its distress to the neighbours. 
Similarly to previous studies \cite{ChenPRL01, GohPRL03, LeePLOS11} 
we set the failure threshold equal to  
the degree $k$ of vertex. The simulation time is discrete and at every step we add a 
grain of sand (distress) on a randomly drawn vertex. 
When the threshold is reached, the vertex topples and distributes a grain of sand 
(unit of distress) to every neighbour. Topplings continue as long as some vertex is above threshold; a single toppling can therefore produce an avalanche. 
The size $s$ of an avalanche is defined as the number of topplings 
occurring until there are no more vertices above threshold.
When the avalanche stops, we add new grains until a new avalanche
starts and so on.
With respect to the original formulation, here we assume that a fraction $P$ of the
vertices are immunized and that such vertices can absorb infinite amount of sand (distress);
this is the analogous to say that a vertex is under the coverage of a central bank. 
To each strategy of choosing the immunized vertices corresponds a policy aimed to limit avalanche propagation, a feature of particularly relevance in the case of a financial crisis;
it is anyhow important to use the least number of immunized vertices to minimize the 
financial cost.

By using this cellular automaton we determine the effects of the assortativity 
on different policies to stop the propagation of distress; in particular, we 
consider both a random and a targeted policy of vertex immunization.

In the random policy we pin $PN_V$ randomly chosen vertices; 
for such a case, assortativity does not play a significant role 
and the fraction $P$ of pinned nodes (i.e. nodes that can absorb any 
amount of distress) is the only control parameter.

For small $P$ avalanche sizes are power-law distributed  with an exponent $\gamma=3/2$ \cite{OlamiPRA88}; this is the worst scenario, where avalanches of the size of the system (systemic crisis) can occur. 

The avalanche size frequency takes the functional form 
$f(s) \propto s^{-3/2} e^{-s/\xi}$; 
the cutoff $\xi$ is a measure of the largest 
possible avalanches and increases with $P$. 
From the point of view of the policy maker, 
it is crucial to limit the size $\xi$ of the systemic crises 
keeping the minimum effort (i.e. small $P$).  

\begin{figure}
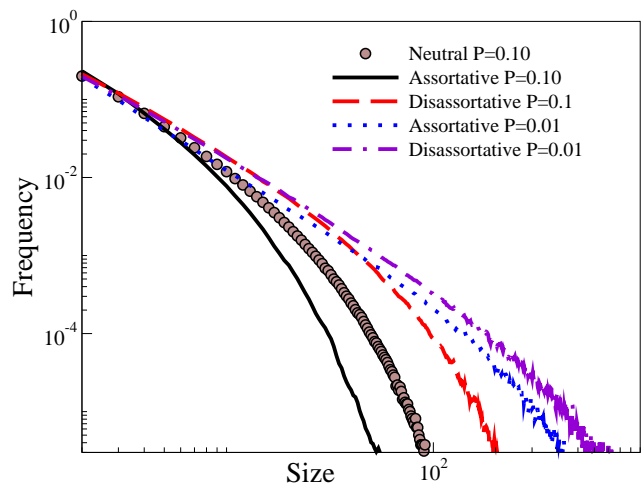

\onefigure[width=0.95\columnwidth]{fig4.eps}
\caption{Avalanche size frequency distribution for an ensemble of $100$
networks of $10000$ nodes each; for each network, we immunise (pin) the $PN_V$ nodes with largest degrees.  
In the figure we compare the most assortative networks with the most disassortative ones.
As expected, increasing $P$ mitigates systemic cascades by depressing the probability 
of the biggest avalanches, i.e. decreasing the cut-off values $\xi$.
At fixed $P$, we observe a clear improvement between the cut-off values $\xi$ for assortative networks with respect to disassortative ones, i.e. the size of the largest avalanche is smaller. 
As an example, immunizing $10\%$ ($P=0.1$) of the banks, 
assortativity improves the effectiveness of the policy by factor $\sim 4$.
Such an effect is monotonous: as an example, the avalanche frequencies at $P=0.10$ 
for neutral ($r \sim 0$) networks has an intermediate behavior respect to
the assortative and the disassortative cases.
}
\label{fig4}
\end{figure}

In the targeted immunization policy we sort the vertices 
according to their degrees and pin the first $PN_V$ starting 
from the largest hub. Such policy takes into account the fragility 
of the hubs in power-law networks \cite{AlbertNAT00}. 
At fixed $P$, we find that targeted immunization is
always more efficient than the random one; moreover, 
the effects of the assortativity become evident even for small values of $P$. 
On Figure \ref{fig4} we present the avalanche size distribution for the targeted 
immunization on networks with large ($J = 10 $) and small ($J = -10 $) assortativity. 
While the exponent of the power law remains $\gamma=3/2$ for both immunization policies,
we find that the value of  the cutoff $\xi$ strongly 
depends on the assortativity of the network. 
In particular, assortative networks are subject to 
much smaller systemic crisis than disassortative ones; 
therefore, assortative networks are easier to immunise.

\section{Conclusions}
In this paper we considered the effects of the topology on the propagation of 
diseases or distress in a network system by means of spectral analysis and simulations.
This problem is often approached by considering the statistical distribution of the 
number of contacts, we instead focused on the two-point degree correlation. 

First, we show that assortativity increases the probability of a pandemic 
(low epidemic threshold) 
while decreasing the speed of diffusive exploration (slow diffusive modes).

Second, we find that in a simple model mimicking financial distress propagation, 
targeted immunization policy may limit the size of financial crisis; we find that assortativity strongly enhances the effectiveness of such a policy. 
These results can be used to devise efficient and fast 
actions to protect infrastructural networks of any kind.

We believe that this paper contributes to a better understanding of 
immunization procedures on complex networks and to a better evaluation 
of the robustness of a given system.
We find that assortativity can not be regarded as a topological intrinsic improvement: 
it enhances time for intervention and improves the effectiveness of financial immunization policies, 
but epidemics/failures are easier to propagate. 
Our findings indicate that policy makers in financial markets need to account carefully for the assortativity of the network to mitigate 
%the spread of bankruptcies 
the spread of economic crises in financial markets.

\acknowledgments
We thank prof. L. Braunstein for pointing out ref.\cite{NohPRE07}.
AS, VZ and GC acknowledge support from FET Open Project "FOC" nr. 255987.
GD acknowledges support by the European project "MOTIA" JLS-2009-CIPS-AG-C1-016.

\bibliographystyle{eplbib}
\bibliography{eplASSORT}

\end{document}